\documentclass[10 pt, conference]{ieeeconf}

\IEEEoverridecommandlockouts 
\makeatletter

\let\proof\@undefined
\let\endproof\@undefined
\let\labelindent\relax
\makeatother

\usepackage{jhoagg}
\usepackage[left=2cm,right=2cm,top=2cm,bottom=1.7cm]{geometry}
\usepackage{cite}

\usepackage{amsmath}
\usepackage{amsfonts}
\usepackage{amssymb,latexsym}
\usepackage[psamsfonts]{eucal}
\usepackage{amsthm}
\usepackage{graphicx}
\usepackage{epsfig}
\usepackage{epstopdf}
\usepackage{epstopdf}
\usepackage{psfrag}
\usepackage{indentfirst}
\usepackage{tikz}
\usepackage{enumitem}
\usepackage{comment}

\usepackage{float}
\usepackage[font=small,skip=0pt]{caption}
\definecolor{fblue}{rgb}{.2,0,.4}
\usepackage[hypertexnames=false,colorlinks,bookmarksopen,bookmarksnumbered,citecolor=fblue,linkcolor=fblue,urlcolor=magenta]{hyperref}

\allowdisplaybreaks

\newtheorem{theorem}{\indent Theorem}

\newtheorem{proposition}{\indent Proposition}

\newtheorem{lemma}{\indent Lemma}

\newtheorem{corollary}{\indent Corollary}

\newtheorem{definition}{\indent Definition}

\newtheorem{example}{\indent Example}

\newcommand\xqed[1]{%
  \leavevmode\unskip\penalty9999 \hbox{}\nobreak\hfill
  \quad\hbox{#1}}
\newcommand\exampletriangle{\xqed{$\triangle$}}

\renewcommand{\(}{\left(}
\renewcommand{\)}{\right)}
\renewcommand{\[}{\left[}
\renewcommand{\]}{\right]}

\title{Multivariable Adaptive Harmonic Steady-State Control for \\ Rejection of Sinusoidal Disturbances Acting on an Unknown System}

\author{\authorblockN{Mohammadreza Kamaldar and Jesse B. Hoagg} 
\authorblockA{Department of Mechanical Engineering, University of Kentucky, Lexington, KY, 40506-0503}}
\AtBeginDocument{%
  \addtolength\abovedisplayskip{-0.45\baselineskip}%
  \addtolength\belowdisplayskip{-0.45\baselineskip}%
}
\allowbreak
\begin{document}
\maketitle
\begin{abstract}
This paper presents an adaptive harmonic steady-state (AHSS) controller, which addresses the problem of rejecting sinusoids with known frequencies that act on a completely unknown multi-input multi-output linear time-invariant system. 
We analyze the stability and closed-loop performance of AHSS for single-input single-output systems.
In this case, we show that AHSS asymptotically rejects disturbances. \footnote{This paper is the preprint version of the following paper: Kamaldar M, Hoagg JB, ``Multivariable adaptive harmonic steady-state control for rejection of sinusoidal disturbances acting on an unknown system," American Control Conference (ACC), 2016. IEEE (pp. 1631-1636), which has been published by IEEE at \href{https://doi.org/10.1109/ACC.2016.7525150}{DOI: 10.1109/ACC.2016.7525150},  and is subject to  IEEE copyright ploicy.}
\end{abstract}

\vspace{-0.5ex}
\section{Introduction}
\vspace{-0.5ex}

The rejection of sinusoidal disturbances is a fundamental control objective in many active noise and vibration control applications such as noise cancellation \cite{elliott1993}, helicopter vibration reduction \cite{friedmann1995}, and active rotor balancing \cite{knospe1996}. 

For an accurately modeled linear time-invariant (LTI) system, the internal-model principle can be used to design a feedback controller capable of rejecting sinusoidal disturbances of known frequencies \cite{davison1976,francis1974,hoagg2008_TAC_IMC}. 
In this case, disturbance rejection is accomplished by incorporating copies of the disturbance dynamics in the feedback loop. 

If, on the other hand, an accurate model of the system is not available, but the open-loop dynamics are asymptotically stable, then adaptive feedforward cancellation can be used to accomplish disturbance rejection \cite{bodson1994,bayard2000}.
One approach for sinusoidal disturbance rejection is harmonic steady-state (HSS) control \cite{elliott1992}, which has been used for helicopter vibration reduction \cite{friedmann1995} and active rotor balancing \cite{knospe1996}.
To discuss HSS control, let $G_{yu}$ denote the control-to-performance transfer function, and assume that there is a single known disturbance frequency $\omega$.
Then, HSS control requires an estimate of $G_{yu}(\jmath \omega)$.
In the SISO case, the estimate of $G_{yu}(\jmath \omega)$, which is a single complex number, must have an angle within $90^\circ$ of $\angle G_{yu}(\jmath \omega)$ to ensure closed-loop stability. 
In the MIMO case, closed-loop stability is ensured provided that the estimate of $G_{yu}(\jmath \omega)$ is sufficiently accurate.
If there are multiple disturbance frequencies, then estimates are required at each frequency. 

For certain applications $G_{yu}(\jmath \omega)$ can be difficult to estimate or subject to change.
To address this uncertainty, online estimation methods have been combined with HSS control \cite{patt2005,chandrasekar2006,pigg2010}. 
For example, a recursive-least-squares identifier is used in \cite{patt2005,chandrasekar2006} to estimate $G_{yu}(\jmath \omega)$ in real time; however, an external excitation signal, which degrades performance, is required to ensure stability.

In this paper, we present a new adaptive harmonic steady-state (AHSS) controller, which is effective for rejecting sinusoids with known frequencies that act on a completely unknown MIMO LTI system. 
We analyze the stability and closed-loop performance for SISO systems.
We show that AHSS asymptotically rejects disturbances.

The new AHSS algorithm in this paper is a frequency-domain method, and all computations are with discrete Fourier transform (DFT) data. 
The AHSS algorithm including DFT is demonstrated on a simulation of an acoustic duct. 

\vspace{-0.5ex}
\section{Notation}
\vspace{-0.5ex}

Let $\BBF$ be either $\BBR$ or $\BBC$.
Let $x_{(i)}$ denote the $i$th element of $x\in\BBF^n$, and let $A_{(i,j)}$ denote the element in row $i$ and column $j$ of $A\in\BBF^{m\times n}$. Let $\|\cdot\|$ be the 2-norm on $\BBF^n$.
Next, let $A^*$ denote the complex conjugate transpose of $A\in\BBF^{m\times n}$, and define $\|A\|_{\rm F}\triangleq\sqrt{{\rm tr~}A^*A}$, which is the Frobenius norm of $A\in\BBF^{m\times n}$.

Let ${\rm spec}(A)\triangleq\{\lambda\in\BBC:\det(\lambda I-A)=0\}$ denote the spectrum of $A\in\BBF^{n \times n}$, and let $\lambda_{\rm max}(A)$ denote the maximum eigenvalue of $A\in\BBF^{n\times n}$, which is Hermitian positive semidefinite. Let $\angle \lambda$ denote the argument of $\lambda\in \BBC$ defined on the interval $(-\pi,\pi]$ rad.
Let ${\rm OLHP}$, ${\rm ORHP}$, and ${\rm CUD}$ denote the open-left-half plane, open-right-half plane, and closed unit disk in $\BBC$, respectively.
Define $\BBN\triangleq\{0,1,2,\cdots\}$ and $\BBZ^{+}\triangleq \BBN\backslash\{0\}$.

\vspace{-0.5ex}
\section{Problem Formulation}
\vspace{-0.5ex}

Consider the system
\begin{align}
\label{Eq:system}
\dot x(t)=Ax(t)+Bu(t)+D_{1}d(t),\\
\label{Eq:Output}
y(t)=Cx(t)+Du(t)+D_{2}d(t),
\end{align} 
where $t\geq 0$, $x(t) \in \BBR ^{n}$ is the state, $x(0)=x_0\in\BBR^n$ is the initial condition, $u(t) \in \BBR^{m}$ is the control, $y(t) \in \BBR^{\ell}$ is the measured performance, $d(t)\in \BBR^{p}$ is the unmeasured disturbance, and $A \in \BBR ^{n\times n}$ is asymptotically stable. Define the transfer functions ${G_{yu}(s)\triangleq C(sI-A)^{-1}B+D,}$ and $G_{yd}(s)\triangleq C(sI-A)^{-1}D_{1}+D_{2}$. Let $\omega_1,\omega_2,\cdots,\omega_q >0$, and consider the tonal disturbance $d(t)=\sum_{i=1}^{q}d_{{\rm c},i}\cos\omega_i t + d_{{\rm s},i} \sin \omega_i t$, 
where $d_{{\rm c},1},\cdots,d_{{\rm c},q},d_{{\rm s},1},\cdots,d_{{\rm s},q}\in \BBR^{p}$.

Our objective is to design a control $u$ that reduces or even eliminates the effect of the disturbance $d$ on the performance $y$. We seek to design a control that relies on no model information of \eqref{Eq:system} and \eqref{Eq:Output}, and requires knowledge of only the disturbance frequencies $\omega_1,\cdots,\omega_q$.

For simplicity, we focus on the case where $d$ is the single-tone disturbance $d(t)=d_{\rm c}\cos\omega  t+d_{\rm s}\sin\omega  t.$ 
However, the adaptive controller presented in this paper generalizes to the case where $d$ consists of multiple tones. We address multiple tones in Example 3.

For the moment, assume that $G_{yu}$, $G_{yd}$, $d_{\rm c}$, and $d_{\rm s}$ are known, and consider the harmonic control $u(t)=u_{\rm c}\cos\omega t+u_{\rm s}\sin\omega t$, where $u_{\rm c}$, $u_{\rm s}$ $\in \BBR^{m}$. Define $\hat u\triangleq u_{\rm c}-\jmath u_{\rm s}$, which is the value at frequency $\omega$ of the DFT obtained from a sampling of $u$. The HSS performance of \eqref{Eq:system} and \eqref{Eq:Output} with control $\hat u$ is 
\begin{align}
\label{Eq:yss1}
y_{\rm hss}&(t,\hat u) \triangleq{\rm Re}\left(M_*\hat u+\hat d\right)\cos\omega t-{\rm Im}\left(M_*\hat u+\hat d\right)\sin\omega t,
\end{align}
where $M_*\triangleq G_{yu}(\jmath\omega )\in\BBC^{\ell\times m}$ and $\hat d\triangleq G_{yd}(\jmath\omega )(d_{\rm c}-\jmath d_{\rm s})\in\BBC^{\ell}$.
The HSS performance $y_{\rm hss}$ is the steady-state response of $y$, that is, $\lim_{t \to \infty}\left[y_{\rm hss}(t,\hat u)-y(t)\right] = 0$ \cite[Chap. 12.12]{bernstein2005}. Consider the cost function
\begin{align}
\label{J_power}
J(\hat u)\triangleq\lim_{t\to\infty}\frac{1}{t}\int\limits_0^t y_{\rm hss}^{\rm T}(\tau,\hat u)y_{\rm hss}(\tau,\hat u)\,{\rm d}\tau,
\end{align}
which is the average power of $y_{\rm hss}$. Define
\begin{align}
\label{y_k+1}
\hat y_{\rm hss}(\hat u)\triangleq M_*\hat u+\hat d,
\end{align}
which is the value at frequency $\omega$ of the DFT obtained from a sampling of $y_{\rm hss}$. It follows from \eqref{Eq:yss1}--\eqref{y_k+1} that $J(\hat u)=\frac{1}{2}\hat y_{\rm hss}^*(\hat u)\hat y_{\rm hss}(\hat u)$. 
The following result provides an expression for an open-loop control $\hat u=u_*$ that minimizes $J$. The proof is omitted due to space limitations.

\textbf{Theorem 1.} Consider the cost function \eqref{J_power}, and assume ${\rm rank~}M_*={\rm min}\{\ell,m\}$. Then, the following statements hold:
\begin{enumerate}
\renewcommand{\theenumi}{{\it \roman{enumi}})}
\renewcommand{\labelenumi}{\theenumi}
\item \label{Thm1Itm1}Assume $\ell> m$, and define $u_{*}\triangleq  -\left(M_*^{*}M_*\right)^{-1}M_*^{*}\hat d$. Then, $\hat y_{\rm hss}(u_*)=\left(I_\ell-M_*(M_*^*M_*)^{-1}M_*^*\right)\hat d$, $J(u_*)=\frac{1}{2}\hat d^*\left(I_\ell-M_*(M_*^*M_*)^{-1}M_*^*\right)\hat d$, and for all $\hat u\in\BBC^{m}\backslash\{u_*\}$, $J(u_*)<J(\hat u)$.
\item \label{Thm1Itm2}Assume $\ell=m$, and define $u_{*}\triangleq -M_*^{-1}\hat d$. Then, $\hat y_{\rm hss}(u_*)=0$, $J(u_*)=0$, and for all $\hat u\in\BBC^{m}\backslash\{u_*\}$, $J(u_*)<J(\hat u)$.
\item \label{Thm1Itm3}Assume $\ell< m$, and let   $u_*\in \{-M_*^*(M_*M_*^*)^{-1}\hat d+(I_m-M_*^*(M_*M_*^*)^{-1}M_*)v:v\in\BBC^m\}$. Then, $\hat y_{\rm hss}(u_*)=0$ and $J(u_*)=0$.
\end{enumerate} 
Theorem 1 provides an expression for a control $u_*$ that minimizes $J$, but $u_*$ requires knowledge of $M_*$ and $\hat d$.

In this paper, we consider a sinusoidal control with frequency $\omega$ but where the amplitude and phase are updated at discrete times. Let $T_{\rm s}>0$ be the update period, and for each $k\in\BBZ^+$, let $u_k\in\BBC^m$ be determined from an adaptive law presented later. Then, for each $k\in\BBN$ and for all $t\in[kT_{\rm s},(k+1)T_{\rm s})$, consider the control
\begin{align}
\label{IDFT}
u(t)=({\rm Re~}u_k)\cos\omega t-({\rm Im~}u_k)\sin\omega t.
\end{align}

Let $y_k\in\BBC^\ell$ denote the value at frequency $\omega$ of the DFT of the sequence obtained by sampling $y$ on the interval $[(k-1)T_{\rm s},kT_{\rm s})$. If $T_{\rm s}$ is sufficiently large relative to the settling time of $G_{yu}$, then $y_{k+1}\approx \hat y_{\rm hss}(u_k)$. For the remainder of this paper, we assume $y_{k+1}=\hat y_{\rm hss}(u_k)$, and it follows from \eqref{y_k+1} that
\begin{align}
\label{Eq:open-loop-MIMO}
y_{k+1}=M_*u_k+\hat d.
\end{align}
In addition, we assume ${\rm rank~}M_*=\min\{\ell,m\}$.

\vspace{-0.5ex}
\section{Harmonic Steady-State Control}
\vspace{-0.5ex}

In this section, we review HSS control, which relies on knowledge of an estimate $M_{\rm e}\in\BBC^{\ell\times m}$ of $M_*$. Let $\rho>0$, and for all $k\in\BBN$, consider the control
\begin{align}
\label{Eq:u-MIMO}
u_{k+1}&= u_{k}-\rho M_{\rm e}^{*}y_{k+1},
\end{align}
where $u_{0}\in\BBC^{m}$ is the initial condition. It follows from \eqref{Eq:open-loop-MIMO} that $y_{k+1}=M_*u_k+\hat d=M_*u_k+y_k-M_*u_{k-1}$, and substituting \eqref{Eq:u-MIMO} yields the closed-loop dynamics
\begin{align}
\label{Eq:closed-loop}
y_{k+1}=y_k-\rho M_*M_{\rm e}^*y_k,
\end{align}
where $k\in\BBZ^+$ and $y_1=M_*u_0+\hat d$. Define $\Lambda\triangleq {\rm spec}(M_{\rm e}^*M_*)\cap{\rm spec}(M_*M_{\rm e}^*)$. The following result presents the stability properties of the closed-loop system \eqref{Eq:closed-loop}. The proof is omitted due to space limitations. 

 \textbf{Theorem 2.} Consider the closed-loop system \eqref{Eq:closed-loop}, which consists of \eqref{Eq:open-loop-MIMO} and \eqref{Eq:u-MIMO}. Assume that $\Lambda\subset {\rm ORHP}$, and assume that $\rho$ satisfies
\begin{align}
\label{mu 2}
0<\rho<\min\limits_{\lambda\in\Lambda}\frac{2{\rm Re~}\lambda}{|\lambda|^2}.
\end{align}
Then, for all $u_0\in\BBC^m$, $u_\infty\triangleq\lim_{k\to\infty}u_k$ exists and $y_\infty\triangleq\lim_{k\to\infty}y_k$ exists.
Furthermore, for all $u_0\in\BBC^m$, the following statements hold:
\begin{enumerate}
\renewcommand{\theenumi}{{\it \roman{enumi}})}
\renewcommand{\labelenumi}{\theenumi}
\item \label{Thm2Itm1} If $\ell> m$, then $u_\infty=-\left(M_{\rm e}^{*}M_*\right)^{-1}M_{\rm e}^*\hat d$ and $y_\infty=[I_{\ell}-M_*(M_{\rm e}^{*}M_*)^{-1}M_{\rm e}^{*}]\hat d$.
\item \label{Thm2Itm2} If $\ell=m$, then $u_\infty=-M_*^{-1}\hat d$ and $y_\infty=0.$ 
\item \label{Thm2Itm3} If $\ell<m$, then $u_\infty=u_0-M_{\rm e}^*(M_*M_{\rm e}^*)^{-1}(M_*u_0+d)$ and  $y_\infty=0.$ 
\end{enumerate} 

Theorem 2 relies on the condition that $\Lambda\subset{\rm ORHP}$. This condition depends on the estimate $M_{\rm e}$ of $M_*$. In the SISO case, $\Lambda\subset{\rm ORHP}$ if and only if $M_{\rm e}$ is within $90^\circ$ of $M_*$, that is, $|\angle (M_{\rm e}/M_*)|<\frac{\pi}{2}$. In this case, \eqref{mu 2} is satisfied by a sufficiently small $\rho>0$.

If $M_{\rm e}=M_*$, then $\Lambda\subset{\rm ORHP}$. In this case, \eqref{mu 2} is satisfied if $\rho<2/\lambda_{\rm max}(M_*^*M_*)$. 

If $\Lambda\cap{\rm OLHP}$ is not empty, then for all $\rho>0$, $I_\ell-\rho M_*M_{\rm e}^*$ has at least one eigenvalue outside the $\rm CUD$. In this case, \eqref{Eq:closed-loop} implies that $y_k$ diverges. 

\vspace{-0.5ex}
\section{Adaptive Harmonic Steady-State Control}
\vspace{-0.5ex}

In this section, we present AHSS control, which does not require any information regarding $M_*$.
Let $\mu\in(0,1]$, $\nu_1>0$, and $u_0\in\BBC^m$, and for all $k\in\BBN$, consider the control
\begin{align}
\label{Eq:u_updateMIMo}
u_{k+1}&= u_{k}-\frac{\mu}{\nu_1+\|M_{k}\|_{\rm F}^2}M_{k}^{*}y_{k+1},
\end{align}
where $M_k\in\BBC^{\ell\times m}$ is an estimate of $M_*$ obtained from the adaptive law presented below.
Note that \eqref{Eq:u_updateMIMo} is reminiscent of the HSS control \eqref{Eq:u-MIMO} except the fixed estimate $M_{\rm e}$ is replaced by the adaptive estimate $M_k$, and the fixed gain $\rho$ is replaced by the $M_k$-dependent gain $\mu/\(\nu_1+\|M_k\|_{\rm F}^2\)$.

To determine the adaptive law for $M_k$, consider the cost function $\SJ:\BBR^{\ell\times m}\times\BBR^{\ell\times m}\to[0,\infty)$ defined by
\begin{align*}
\mathcal{J}(M_{\rm r},M_{\rm i})& \triangleq \frac{1}{2}\left \|  (M_{\rm r}+\jmath M_{\rm i})( u_{k} - u_{k-1} ) - (y_{k+1} - y_{k} ) \right \|^2.
\end{align*}
Note that $\SJ({\rm Re~}M_*,{\rm Im~}M_*)=0$, that is, $M_*$ minimizes $\SJ$. Define the complex gradient
\begin{align}
\label{Eq:gradJ}
\nabla \SJ(M_{\rm r},M_{\rm i}) & \triangleq \frac{\partial\SJ( M_{\rm r},M_{\rm i})}{\partial M_{\rm r}}+\jmath \frac{\partial\SJ( M_{\rm r}, M_{\rm i})}{\partial M_{\rm i}}\nonumber\\
&=\left [ (M_{\rm r}+\jmath M_{\rm i}) ( u_{k} - u_{k-1} ) - (y_{k+1} - y_{k} ) \right ] \nonumber\\
&\qquad\times( u_{k} - u_{k-1})^*,
\end{align}
which is the direction of the maximum rate of change of $\SJ$ with respect to $ M_{\rm r}+\jmath M_{\rm i} $ \cite{brandwood1983}. Let $M_0\in\BBC^{\ell\times m}\backslash\{0\}$, $\gamma\in(0,1]$, and $\nu_2>0$, and for all $k\in\BBZ^{+}$, consider the adaptive law
\begin{align}
\label{AdaptiveLaw}
M_{k}&\triangleq M_{k-1} - \eta_{k}\nabla \SJ({\rm Re~}M_{k-1},{\rm Im~}M_{k-1}),
\end{align}
where
\begin{align}
\label{xi_k}
\eta_k\triangleq\frac{\gamma (\nu_1+\|M_{k-1}\|_{\rm F}^2)^2}{\nu_2\mu^2+(\nu_1+\|M_{k-1}\|_{\rm F}^2)^2 \left \| u_{k}-u_{k-1} \right \|^2}.
\end{align}
Using \eqref{Eq:gradJ}--\eqref{xi_k}, it follows that, for all $k\in\BBN$, 
\begin{align}
\label{Eq:c_updateMIMO}
&M_{k}=M_{k-1}-\eta_k \Big[M_{k-1}(u_k-u_{k-1})-(y_{k+1}-y_{k})\Big]\nonumber\\
&\qquad\times(u_k-u_{k-1})^*.
\end{align}

\begin{figure}[t]
	\centering
	\includegraphics[scale=1]{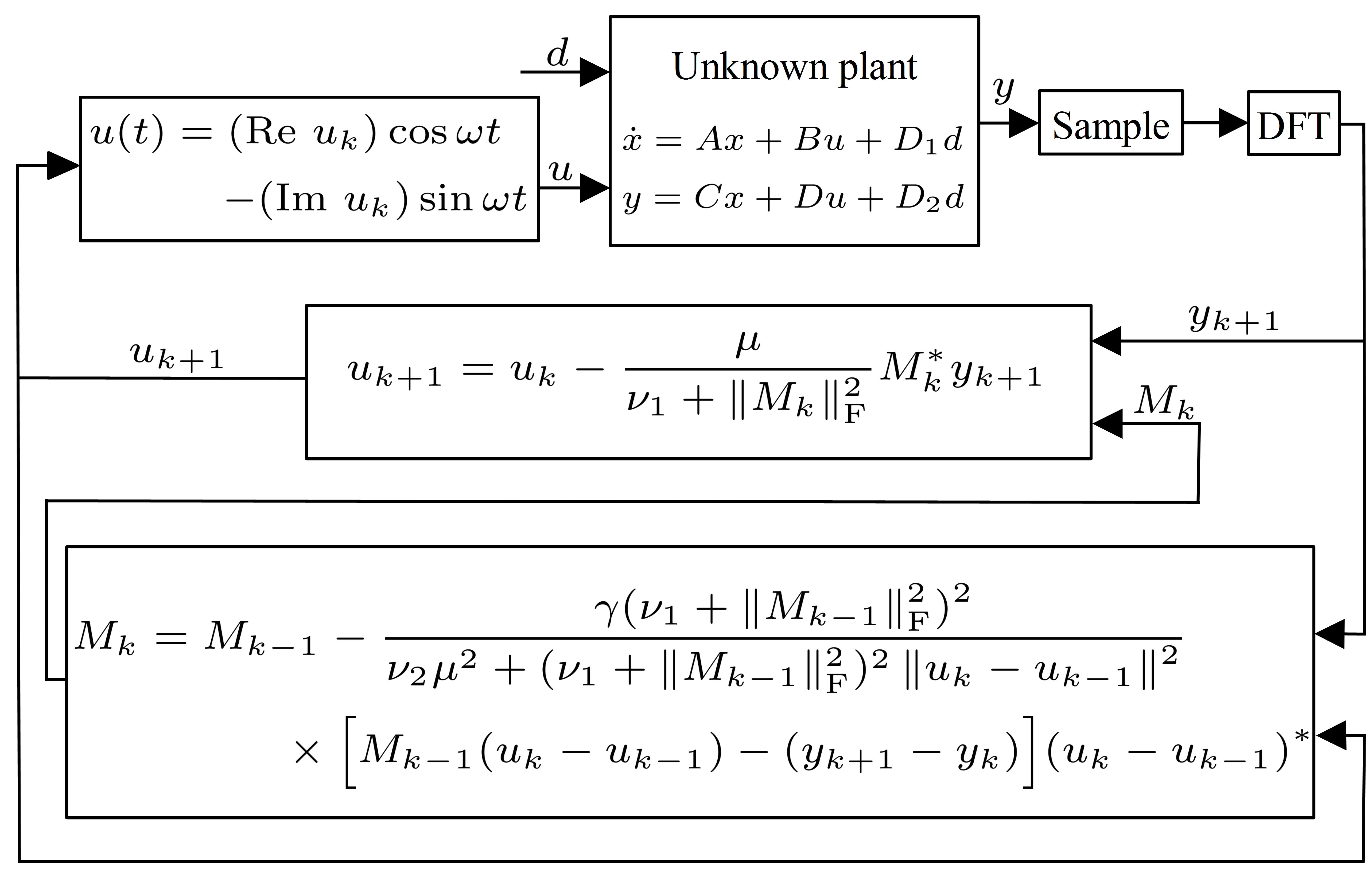}
	\caption{AHSS control architecture.}\label{fig:control architecture}
	\vspace{-2.5ex}
\end{figure}

Thus, the AHSS control is given by \eqref{Eq:u_updateMIMo}, \eqref{xi_k}, and \eqref{Eq:c_updateMIMO}. The control architecture is shown in Fig. \ref{fig:control architecture}. All AHSS computations are performed using complex DFT signals. At time $kT_{\rm s}$, the control $u$ is updated using \eqref{IDFT} and the complex signal $u_k$. Note that $u_k$ is calculated using $y_k$, which is the DFT of $y$ at frequency $\omega$ sampled over the interval $[(k-1)T_{\rm s},kT_{\rm s})$, which corresponds to the time between the $k-1$ and $k$ steps.

The update period $T_{\rm s}$ must be sufficiently large such that the harmonic steady-state assumption $y_{k+1}\approx \hat y_{\rm hss}(u_k)$ is valid. Numerical testing suggests that $T_{\rm s}$ should be at least as large as the settling time associated with the slowest mode of $A$, that is, $T_{\rm s}>4/(\zeta\omega_{\rm n})$, where $\zeta$ and $\omega_{\rm n}$ are the damping ratio and natural frequency of the slowest mode of $A$.

The AHSS controller parameters are $\mu\in(0,1]$, $\gamma\in(0,1]$, $\nu_1>0$, and $\nu_2>0$. The gains $\mu$ and $\gamma$ influence the step size of the $u_k$ and $M_k$ update equations, respectively. The gain $\nu_1$ and $\nu_2$ influence the normalization of the $u_k$ and $M_k$ update equations, respectively. 

\vspace{-0.5ex}
\section{Stability Analysis}
\vspace{-0.5ex}

The following result provides stability properties of the estimator \eqref{Eq:c_updateMIMO}. The proof follows from direct computation and is omitted due to space limitation.

\textbf{Proposition 1.} Consider the open-loop system \eqref{Eq:open-loop-MIMO}, and the AHSS control \eqref{Eq:u_updateMIMo}, \eqref{xi_k}, and \eqref{Eq:c_updateMIMO}, where $\mu\in(0,1]$, $\gamma\in(0,1]$, $\nu_1>0$, and $\nu_2>0$. Then, for all $u_0\in\BBC$ and $M_0\in\BBC\backslash\{0\}$, the estimate $M_k$ is bounded, and for all $k\in\BBZ^+$, 
\begin{align}
\label{prop1}
\|M_{k}-M_*\|_{\rm F}^2&-\|M_{k-1}-M_*\|_{\rm F}^2\nn\\
&\leq-\dfrac{\gamma \| (M_{k-1}-M_*) M_{k-1}^{*}y_{k}\|^2}{\nu_2+\|M_{k-1}^{*}y_{k}\|^2}.
\end{align}

Proposition 1 implies that $\|M_k-M_*\|_{\rm F}^2$ is nonincreasing. 

We now analyze closed-loop performance under the assumption that the open-loop system is SISO. 
Define $u_*\triangleq -\hat d/M_*$, which exists because $M_*\neq 0$. Note that if $u_k\equiv u_*$, then $y_k\equiv 0$. 
Next, \eqref{Eq:open-loop-MIMO} implies that $y_{k+1}=M_*u_k+\hat d=M_*u_k+y_k-M_*u_{k-1}$, and substituting \eqref{Eq:u_updateMIMo} yields
\begin{align}
\label{Eq:closed-loop-siso}
y_{k+1}=y_k-\frac{\mu M_*M_{k-1}^*y_k}{\nu_1+|M_{k-1}|^2},
\end{align}
where $k\in\BBZ^+$ and $y_1=M_*u_0+\hat d$. Furthermore, \eqref{xi_k} and \eqref{Eq:c_updateMIMO} can be written as  
\begin{align}
\label{Eq:c_updateSISO}
M_{k}&=M_{k-1}-\frac{\gamma}{\nu_2+|M_{k-1}|^2|y_{k}|^2}\bigg[M_{k-1}M_{k-1}^*y_{k}\nn\\
&\qquad+\frac{\nu_1+|M_{k-1}|^2}{\mu}(y_{k+1}-y_{k})\bigg]y_{k}^*M_{k-1}.
\end{align}
Define $\SM\triangleq \BBC\backslash\{x\in\BBC:|\angle x-\angle M_*|=\pi\}$, which is the set of all complex numbers except those numbers that are exactly $180^\circ$ from $M_*$. 
The following result provides the closed-loop SISO stability properties. 
The proof is in Appendix A.

\textbf{Theorem 3.} Consider the closed-loop system \eqref{Eq:closed-loop-siso} and \eqref{Eq:c_updateSISO}, which consists of \eqref{Eq:open-loop-MIMO}, \eqref{Eq:u_updateMIMo}, \eqref{xi_k}, and \eqref{Eq:c_updateMIMO}, where $\mu\in(0,1]$, $\gamma\in(0,1]$, $\nu_1>0$, and $\nu_2>0$. Then, $(y_{k},M_{k})\equiv(0,M_*)$ is a Lyapunov stable equilibrium of \eqref{Eq:closed-loop-siso} and \eqref{Eq:c_updateSISO}. Furthermore, for all initial conditions $u_0\in\BBC$ and $M_0\in\SM\backslash\{0\}$, $M_k$ is bounded, $\lim_{k\to\infty} u_{k}=u_{*}$ and $\lim_{k\to\infty} y_{k}=0.$

\vspace{-0.5ex}
\section{Numerical Examples}
\vspace{-0.5ex}

\begin{figure}[h]
\begin{tikzpicture}
\definecolor{fblue}{rgb}{.2,0.6,0.99}
\filldraw[line width=0.15mm,fill=white!20!white,draw=black!50!black] (-4.2,.75) -- (4.2,.75).. controls (4.3,0) .. (4.2,-.75)--(-4.2,-.75);\filldraw[thin,fill=white!20!white,draw=black!50!black] (-4.2,0) ellipse [x radius=.1, y radius=.75];
\filldraw[thin,fill=white!50!black,draw=black!50!black] (-4.2,0) ellipse [x radius=.03, y radius=.60];
\draw[very thick,green!50!black](-3.14,.30) node {$\phi_{1}$};
\filldraw [fill=green!20!white, draw=green!50!black] (-3.24,-0.1) rectangle (-3.04,.1);
\draw[very thin,green!50!black,font=\small](-3.14,-.30) node {Feedback};
\draw[very thin,green!50!black,font=\small](-3.14,-.60) node {microphone};
\draw[very thick,green!50!black](2.94,.30) node {$\phi_{2}$};
\draw[very thin,green!50!black,font=\small](2.94,-.30) node {Feedback};
\draw[very thin,green!50!black,font=\small](2.94,-.60) node {microphone};
\filldraw [fill=green!20!white, draw=green!50!black] (3.04,-0.1) rectangle (2.84,.1);
\draw[very thick,red!50!black](-.21,1.35) node 
{$d$};
\filldraw [fill=red!20!white, draw=red!50!black] (-0.71,.75) -- (.29,.75) -- (.14,.95) -- (-.56,.95) -- (-0.71,.75);\filldraw [fill=red!20!white, draw=red!50!black] (-.56,.95) rectangle (.14,1.15);
\draw[very thin,red!50!black,font=\small](-.21,1.95) node {Disturbance};
\draw[very thin,red!50!black,font=\small](-.21,1.65) node {speaker};
\draw[very thick,blue!50!black](-2.52,1.35) node {$\psi_{1}$};
\filldraw [fill=fblue, draw=blue!50!black] (-3.02,.75) -- (-2.02,.75) -- (-2.17,.95) -- (-2.87,.95) -- (-3.02,.75);\filldraw [fill=fblue, draw=blue!50!black] (-2.87,.95) rectangle (-2.17,1.15);
\draw[very thin,blue!50!black,font=\small](-2.52,1.95) node {Control};
\draw[very thin,blue!50!black,font=\small](-2.52,1.65) node {speaker};
\draw[very thick,blue!50!black](1.2,1.35) node {$\psi_{2}$};
\filldraw [fill=fblue, draw=blue!50!black] (1.7,.75) -- (.7,.75) -- (.85,.95) -- (1.55,.95) -- (1.7,.75);\filldraw [fill=fblue, draw=blue!50!black] (1.55,.95) rectangle (0.85,1.15);
\draw[very thin,blue!50!black,font=\small](1.2,1.95) node {Control};
\draw[very thin,blue!50!black,font=\small](1.2,1.65) node {speaker};
\draw[very thin,gray] (-4.2,-.75) -- (-4.2,-3.30);
\draw[thin,gray][<->] (-4.2,-1.20) -- (-3.14,-1.20);
\draw[very thin,black](-3.67,-1.05) node {$\xi_{\phi_{1}}$};
\draw[very thin,gray] (-3.14,-.75) -- (-3.14,-1.25);

\draw[thin,gray][<->] (-4.2,-1.65) -- (-2.52,-1.65);
\draw[very thin,black](-3.36,-1.50) node {$\xi_{\psi_{1}}$};
\draw[very thin,gray] (-2.52,-.75) -- (-2.52,-1.70);

\draw[very thin,gray] (-.21,-.75) -- (-.21,-2.1);
\draw[thin,gray][<->] (-4.2,-2.05) -- (-.21,-2.05);
\draw[very thin,black](-2.1,-1.90) node {$\xi_{d}$};

\draw[very thin,gray] (1.2,-.75) -- (1.2,-2.5);
\draw[thin,gray][<->] (-4.2,-2.45) -- (1.2,-2.45);
\draw[very thin,black](-1.5,-2.30) node {$\xi_{\psi_2}$};

\draw[very thin,gray] (2.94,-.75) -- (2.94,-2.9);
\draw[thin,gray][<->] (-4.2,-2.85) -- (2.94,-2.85);
\draw[very thin,black](-.63,-2.70) node {$\xi_{\phi_2}$};

\draw[very thin,gray] (4.2,-.75) -- (4.2,-3.3);
\draw[thin,gray][<->] (-4.2,-3.25) -- (4.2,-3.25);
\draw[very thin,black](0,-3.1) node {$L$};
\end{tikzpicture}
\caption{Acoustic duct.}
\label{fig:Acoustic duct}
\end{figure}
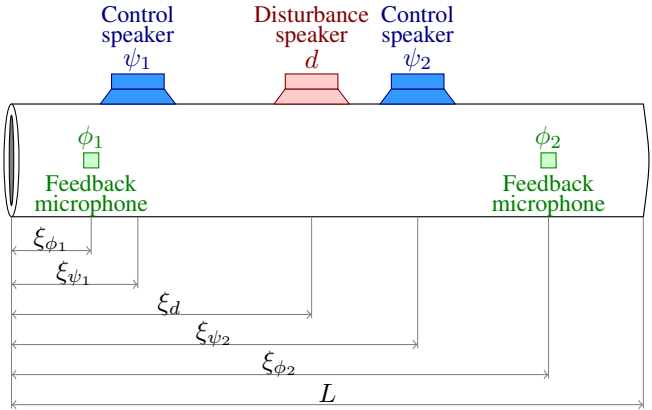

Consider the acoustic duct of length $L=2~{\rm m}$  shown in Fig. \ref{fig:Acoustic duct}, where all measurements are from the left end of the duct. A disturbance speaker is at  $\xi_d=0.95$ m, while 2 control speakers are at  $\xi_{\psi_1}=0.4$ m and $\xi_{\psi_2}=1.25$ m. 
All speakers have cross-sectional area $A_{\rm s}=0.0025{~\rm m}^2$.
The equation for the acoustic duct is 
\begin{align*}
\frac{1}{c^2}\frac{\partial^2 p(\xi,t)}{\partial t^2}&=\frac{\partial^2 p(\xi,t)}{\partial \xi^2}+\rho_0\dot \psi_1(t)\delta(\xi-\xi_{\psi_1})\nonumber\\
&\qquad+\rho_0\dot \psi_{2}(t)\delta(\xi-\xi_{\psi_2})+\rho_0\dot d(t)\delta(\xi-\xi_d),
\end{align*}
where $p(\xi,t)$ is the acoustic pressure, $\delta$ is the Dirac delta, $c=343$ m/s is the phase speed of the acoustic wave, $\psi_{{1}}$ and $\psi_{{2}}$ are the speaker cone velocities of the control speakers, $d$ is the speaker cone velocity of the disturbance speaker, and $\rho_0=1.21$ kg/${\rm m}^2$ is the equilibrium density of air at room conditions. See \cite{hong1996} for more details.

Using separation of variables and retaining $r$ modes, the solution $p(\xi,t)$ is approximated by $p(\xi,t)=\sum_{i=0}^r q_i(t)V_i(\xi)$, where for $i=1,\cdots,r$, $V_i(\xi)\triangleq  c\sqrt{2/L}\sin i\pi\xi/L$, and $q_i$ satisfies the differential equation \eqref{Eq:system}, where
\begin{gather*}
{\small x(t)=\[\begin{array}{ccccc}\int_0^t q_1(\sigma){\rm d}\sigma & q_1(t)& \cdots & \int_0^t q_r(\sigma){\rm d}\sigma & q_r(t)\end{array}\]^{\rm T}},\\
{\tiny A= {\rm diag}  \(\[\begin{array}{cc}0&1\\-\omega_{{\rm n}_{1}}^2&-2\zeta_1\omega_{{\rm n}_{1}}\end{array}\]\right.,\cdots,\left.\[\begin{array}{cc}0&1\\-\omega_{{\rm n}_{r}}^2&-2\zeta_r\omega_{{\rm n}_{r}}\end{array}\]\)},\\
B= \frac{\rho_0}{A_{\rm s}}\[\begin{array}{ccccc}0&V_1(\xi_{\psi_{1}})&\cdots&0& V_r(\xi_{\psi_{1}})\\0&V_1(\xi_{\psi_{2}})&\cdots&0& V_r(\xi_{\psi_{2}})\end{array}\]^{\rm T},\\
D_1=  \frac{\rho_0}{A_{\rm s}}\[\begin{array}{ccccc}0&V_1(\xi_{d})&\cdots&0& V_r(\xi_{d})\end{array}\]^{\rm T},
\end{gather*}
and for $i=1,\cdots,r$, $\omega_{{\rm n}_i}\triangleq i\pi c/L$ is the natural frequency of the $i$th mode, and $\zeta_i=0.2$ is the assumed damping ratio of the $i$th mode.

Two feedback microphones are in the duct at $\xi_{\phi_1}=0.3$ m and $\xi_{\phi_2}=1.7$ m, and they measure the acoustic pressures $\phi_1(t)=p(\xi_{\phi_1},t)$ and $\phi_2(t)=p(\xi_{\phi_2},t)$, respectively. Thus, for $i=1,2$, $\phi_i(t)=C_ix(t)$, where $
C_i=  \frac{\rho_0}{A_{\rm s}}[\begin{array}{ccccc}0&V_1(\xi_{\phi_{i}})&\cdots&0& V_r(\xi_{\phi_{i}})\end{array}].$
For all examples, $r=5$ and $x(0)=0$. The DFT is performed using a 1 kHz sampling frequency. The HSS and AHSS parameters are $T_{\rm s}=0.1$ s, $u_0=0$, $\mu=\gamma=0.2$, $\nu_1=\nu_2=0.1\|M_0\|_{\rm F}^{2}$, $\rho=\mu/(\nu_1+\|M_0\|_{\rm F}^2)$, and $M_{\rm e}=M_0$, where $M_0$ is specified in each example. The following examples consider the acoustic duct with different control speaker and feedback microphone configurations. 
Let $\omega_1=251~{\rm rad/s}$ and $\omega_2=628~{\rm rad/s}$.

\begin{figure}[t!]
\centering
\includegraphics[scale=.97]{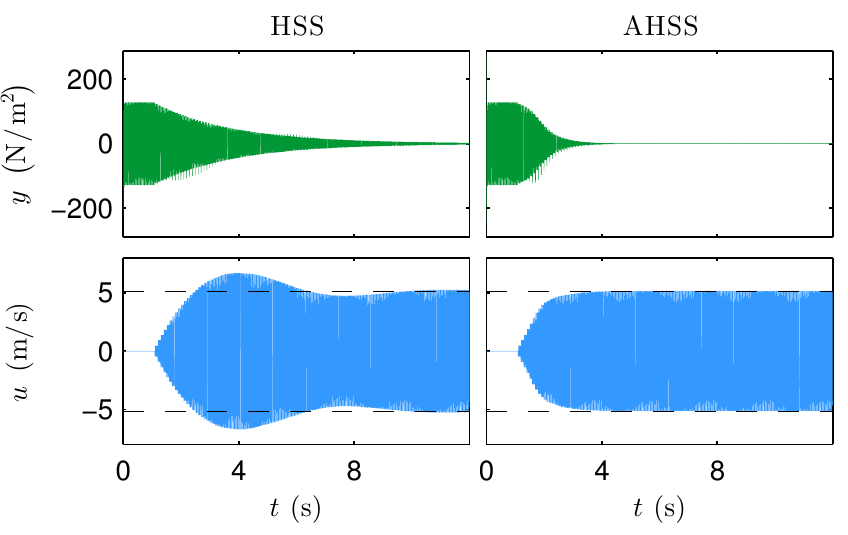}
\caption{For a SISO plant with $|\angle M_0-\angle M_*|<90^\circ$, both HSS and AHSS yield $y(t)\to 0$ as $t\to\infty$. Dashed lines show $\pm|u_*|$.}
\label{fig:Ex1-2}
\vspace{-.3cm}
\end{figure}

\begin{figure}[t!]
\includegraphics[scale=.97]{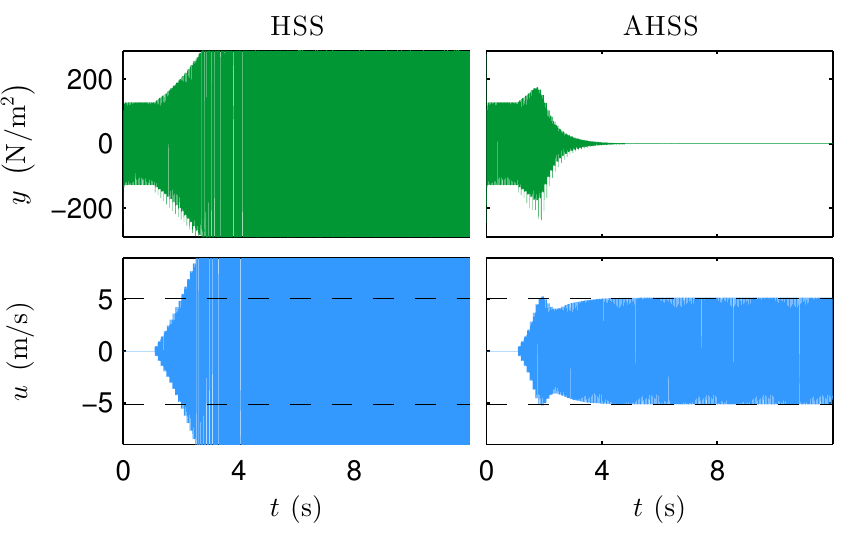}
\caption{For a SISO plant with $|\angle M_0-\angle M_*|>90^\circ$, the response $y$ with HSS diverges, whereas AHSS yields $y(t)\to 0$ as $t\to\infty$. Dashed lines show $\pm|u_*|$.}
\label{fig:Ex1-3}
\vspace{-.5cm}
\end{figure}

\begin{figure}[t!]
\centering
\includegraphics[scale=.97]{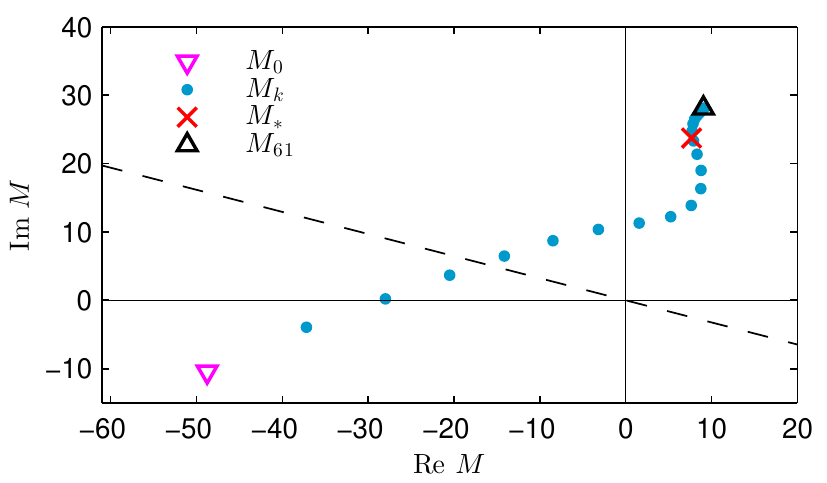}
\caption{Trajectory of $M_k$ with AHSS for a SISO plant where $|\angle(M_0/M_*)|>\frac{\pi}{2}$. The dashed line shows the locus of $M$ such that $|\angle(M/M_*)|=\frac{\pi}{2}$, which is HSS stability boundary for $M_{\rm e}$. Selection of $M_{\rm e}=M_0$ from the lower region, where $|\angle(M/M_*)|>\frac{\pi}{2}$, results in an unstable response with HSS, whereas AHSS yields asymptotic disturbance rejection for all $M_0\in\SM$.}
\label{fig:Ex1-4}
\vspace{-.5cm}
\end{figure}

\textbf{Example 1. } \textit{SISO} $(m=\ell=1)$. 
Let $u=\psi_{{1}}$, $y=\phi_1$, $\psi_{{2}}=0$, and $d=\sin\omega_1 t+2\cos\omega_1t$. First, consider the case where $M_0$ is within $90^\circ$ of the $M_*$, specifically, $M_0=2e^{\jmath\frac{\pi}{3}}M_*$. Figure \ref{fig:Ex1-2} shows $y$ and $u$ for HSS and AHSS. The control is turned on after 1 s. Both HSS and AHSS yield asymptotic disturbance rejection. Next, let $M_0=2e^{\jmath\frac{2\pi}{3}}M_*$ which is not within $90^\circ$ of $M_*$. Figure \ref{fig:Ex1-3} shows  $y$ and $u$ for HSS and AHSS. In this case, $y$ with HSS diverges, whereas $y$ with AHSS converges to zero. Figure \ref{fig:Ex1-4} shows the trajectory of the estimate $M_k$, which moves toward $M_*$. Proposition 1 states that $|M_k-M_*|$ is nondecreasing; however, this result assumes that $y$ reaches harmonic steady state. Figure \ref{fig:Ex1-4} shows that $M_k-M_*$ may increase slightly in practice but generally decreases.
\exampletriangle

\begin{figure}[h]
\centering
\includegraphics[scale=.97]{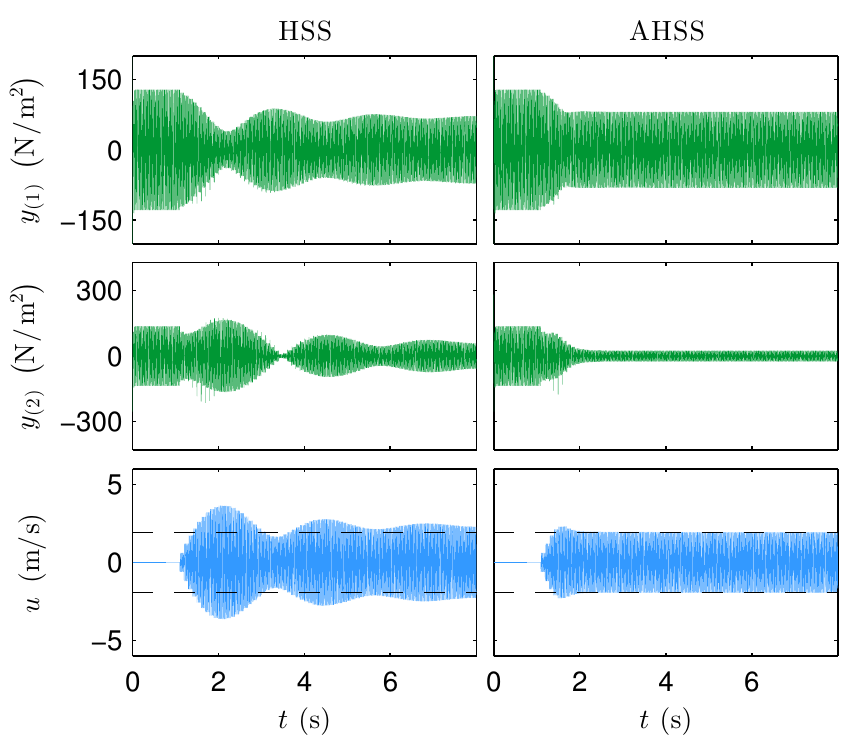}
\caption{For a single-input two-output plant satisfying \eqref{mu 2}, both HSS and AHSS minimize $\lim_{k\to\infty}\|y_k\|$. Dashed lines show ${\pm|u_*|.}$}.
\label{fig:Ex3-1}
\vspace{-.5cm}
\end{figure}

\textbf{Example 2. } \textit{Single-input two-output} $(m=1~and~\ell=2)$.
Let $u=\psi_{{1}}$, $y=[\begin{array}{cc}
\phi_1&\phi_2\end{array}]^{\rm T}$, $\psi_{{2}}=0$, and $d=\sin\omega_1 t+2\cos\omega_1t$. First, consider the case where $M_0$ is selected such that \eqref{mu 2} is satisfied, specifically, $M_0=[\begin{array}{cc}1.5e^{\jmath \frac{\pi}{4}}\(M_{*}\)_{(1,1)}&0.5e^{\jmath \frac{\pi}{3}}\(M_*\)_{(2,1)}\end{array}]^{\rm T}$. Note that the optimal control is $u_*=-1.66+\jmath 0.98$, which minimizes the average power \eqref{J_power}.
Figure \ref{fig:Ex3-1} shows $y$ and $u$ for HSS and AHSS. The control is turned on after 1 s. In this case, HSS and AHSS each yield $u_k\to u_*$ as $k\to\infty$. Thus, $\lim_{k\to\infty}\|y_k\|$ is minimized. Next, let  $M_0=[\begin{array}{cc}1.5e^{\jmath \frac{3\pi}{4}}\(M_{*}\)_{(1,1)}&0.5e^{\jmath \frac{2\pi}{3}}\(M_{*}\)_{(2,1)}\end{array}]^{\rm T}$, which does not satisfy \eqref{mu 2}. Figure \ref{fig:Ex3-1} shows $y$ and $u$ for HSS and AHSS. In this case, $y$ with HSS diverges, whereas $y$ with AHSS converges and $u_k\to u_*$ as $k\to\infty$, which implies that $\lim_{k\to\infty}\|y_k\|$ is minimized.
\exampletriangle


\begin{figure}[h]
\centering
\includegraphics[scale=.97]{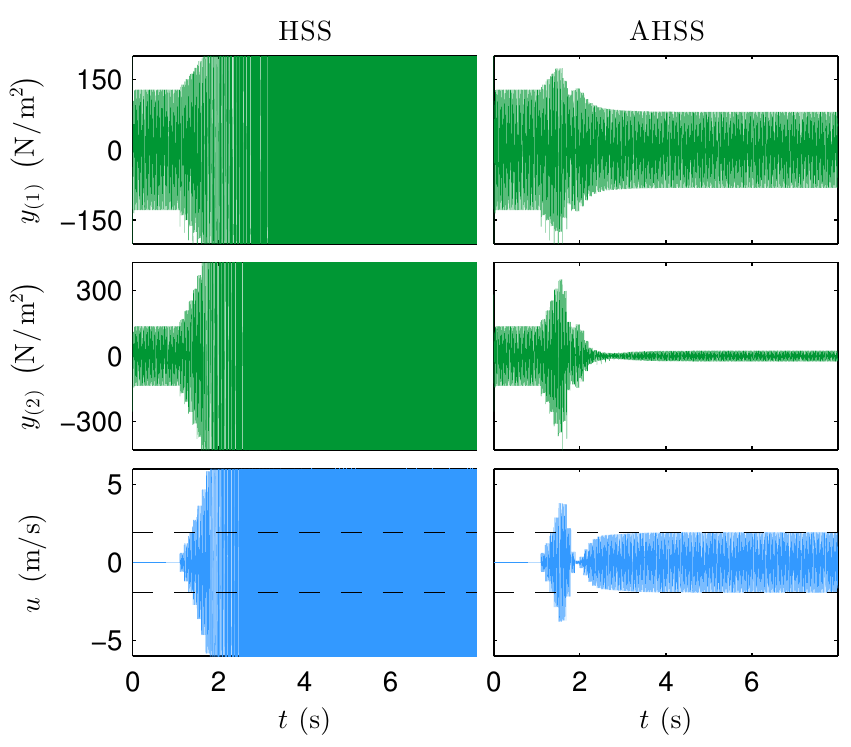}
\caption{For a single-input two-output plant that does not satisfy \eqref{mu 2}, the response $y$ with HSS diverges, whereas AHSS minimizes $\lim_{k\to\infty}\|y_k\|$. Dashed lines show $\pm|u_*|$.}
\label{fig:Ex3-2}
\vspace{-.5cm}
\end{figure}

\begin{figure}[h]
\centering
\includegraphics[scale=.97]{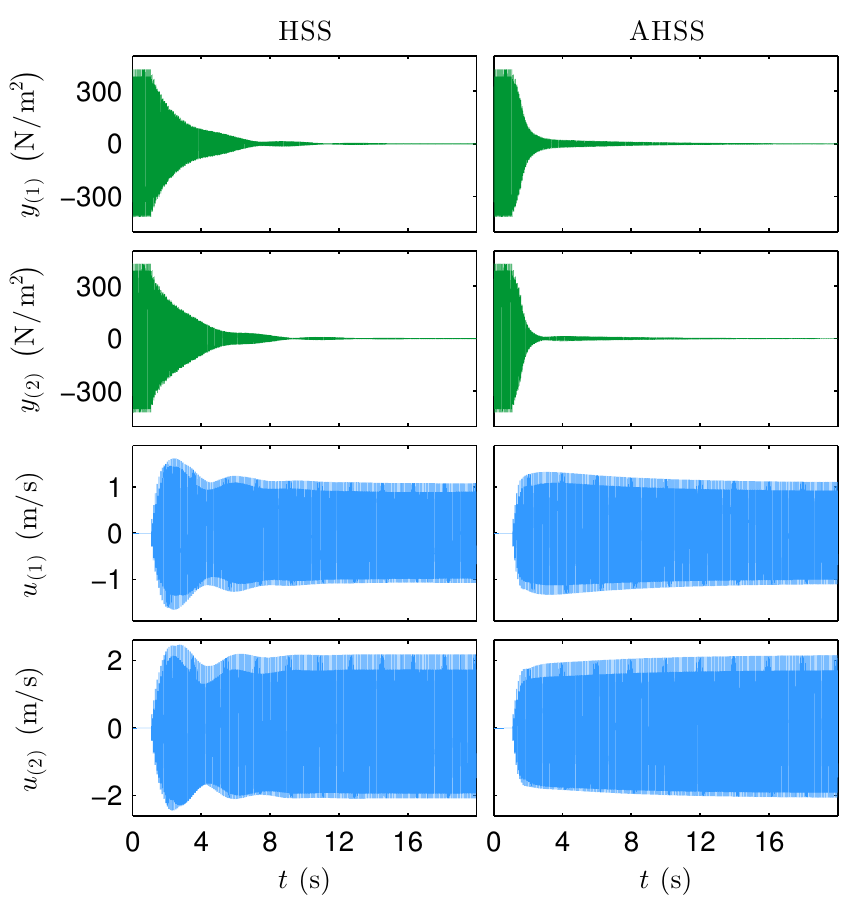}
\caption{For a MIMO plant that satisfies \eqref{mu 2} with a 2-tone disturbance, both HSS and AHSS yield $y(t)\to 0$ as $t\to\infty$.}
\label{fig:Ex4-1}
\vspace{-.5cm}
\end{figure}

\begin{figure}[h!]
\centering
\includegraphics[scale=.97]{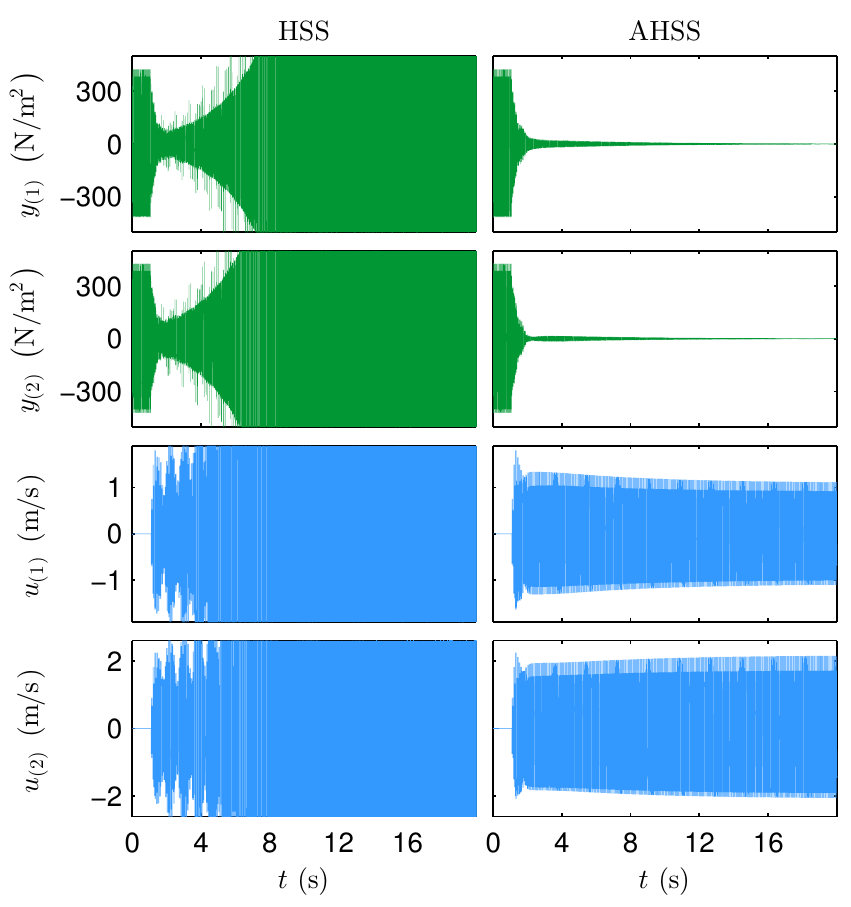}
\caption{For a MIMO plant that does not satisfy \eqref{mu 2} with a 2-tone disturbance, the response $y$ with HSS diverges, whereas AHSS yields $y(t)\to 0$ as $t\to\infty$.}
\label{fig:Ex4-2}
\vspace{-.75cm}
\end{figure}

\textbf{Example 3. }\textit{MIMO ($m=2~and~\ell=2$) with a two-tone disturbance}. 
Let $u=[\begin{array}{cc}\psi_1 & \psi_{{2}} \end{array}]^{\rm T}$, $y=[\begin{array}{cc}
\phi_1&\phi_2\end{array}]^{\rm T}$, and $d=\sin\omega_1 t+\sin\omega_2 t+\cos\omega_1t+\cos\omega_2 t$, which is a two-tone disturbance. Define $M_{*,1}\triangleq G_{yu}(\jmath \omega_1)$, and $M_{*,2}\triangleq G_{yu}(\jmath \omega_2)$. Since $d$ has 2 tones, we use 2 copies of the HSS or AHSS algorithm---one copy at each disturbance frequency.
Let $M_{\rm 1,0}$ and $M_{\rm 2,0}$ denote the initial estimates of $M_{*,1}$ and $M_{*,2}$.
First, consider the case where $M_{1,0}$ and $M_{2,0}$ are such that \eqref{mu 2} is satisfied, specifically, $M_{1,0}=0.6e^{\jmath\frac{\pi}{6}}M_{*,1}$, and $M_{2,0}=0.9e^{\jmath\frac{\pi}{3}}M_{*,2}$.
Figure \ref{fig:Ex4-1} shows $y$ and $u$ for HSS and AHSS. The control is turned on after 1 s. Both HSS and AHSS yield asymptotic disturbance rejection. Next, consider the case where $M_{1,0}=0.2e^{\jmath\frac{\pi}{7}}M_{*,1}$, and $M_{2,0}=0.6e^{\jmath\frac{\pi}{14}}M_{*,2}$, which do not satisfy \eqref{mu 2}. Figure \ref{fig:Ex4-2} shows $y$ and $u$ for HSS and AHSS. In this case, $y$ with HSS diverges, whereas $y$ with AHSS converges to zero.   
\exampletriangle


\section*{Appendix A: Proof of Theorem 3}

\begin{proof}[\indent Proof]
Define $\tilde M_{k}\triangleq M_{k}-M_*$, $V_{M}(\tilde M_{k})\triangleq |\tilde M_{k}|^{2},$ and $\Delta V_{M}(k)\triangleq V_{M}(\tilde M_{k})-V_{M}(\tilde M_{k-1}).$
It follows from Proposition 1 that for all $k\in\BBZ^+$
\begin{align}
\label{Eq:deltaVM}
\Delta V_{M}(k)&\leq -\frac{\gamma|M_{k-1}|^2|y_{k}|^2|\tilde M_{k-1}|^2}{\nu_2+|M_{k-1}|^2|y_{k}|^2}.
\end{align}
Next, define $V_{y}(y_k) \triangleq |y_k|^2 $ and $\Delta V_{y}(k)\triangleq V_y(y_{k+1})-V_y(y_{k})$. Evaluating $\Delta V_y(k)$ along the trajectories of \eqref{Eq:closed-loop-siso} yields
\begin{align}
\label{Delta J0}
\Delta V_y(k)&=-\frac{\mu|y_{k}|^{2}}{\nu_1+|M_{k-1}|^2}\bigg(2{\rm Re~}M_*  M_{k-1}^*\nonumber\\
&\qquad-\frac{\mu|M_*|^{2}|M_{k-1}|^{2}}{\nu_1+|M_{k-1}|^2} \bigg),
\end{align}
Note that $|\tilde M_{k-1}|^{2}=|M_{k-1}|^{2}+|M_*|^{2}-2{\rm Re~}M_*  M_{k-1}^*$, 
and it follows from \eqref{Delta J0} that
\begin{align}
\label{Delta J}
\Delta V_y(k)&=-\frac{\mu|y_{k}|^{2}}{\nu_1+|M_{k-1}|^2}\left(|M_{k-1}|^{2}+|M_*|^{2}-|\tilde M_{k-1}|^{2}\right.\nonumber\\
&\qquad\left.-\frac{\mu|M_*|^{2}|M_{k-1}|^{2}}{\nu_1+|M_{k-1}|^2}\right).
\end{align}
Define the Lyapunov function $V(y_{k},\tilde M_{k-1})\triangleq \ln(1+aV_y(y_{k}))+bV_{M}(\tilde M_{k-1})$,
where $a\triangleq (|M_{0}|+2|M_*|)^2/\nu_2$, and $b>0$ is provided later. Consider the Lyapunov difference
\begin{align}
\label{DeltaVmaa}
\Delta V(k)&\triangleq V(y_{k+1},\tilde M_{k})-V(y_{k},\tilde M_{k-1}).
\end{align}

Since for all $x>0$, $\ln x\leq x-1$, evaluating $\Delta V$ along the trajectories of \eqref{Eq:closed-loop-siso} and \eqref{Eq:c_updateSISO} yields
\begin{align}
\label{Eq:Ineq12}
\Delta V(k)&=\ln\left(1+\frac{a\Delta V_y(k)}{1+aV_y(y_{k})}\right)+b\Delta V_{M}(k)\nonumber\\
&\leq \frac{a\Delta V_y(k)}{1+aV_y(y_k)}+b\Delta V_M(k).
\end{align}
Substituting \eqref{Eq:deltaVM} and \eqref{Delta J} into \eqref{Eq:Ineq12} yields
\begin{align}
\label{Eq:ineq20}
\Delta V(k)&\leq-\frac{a\mu|y_{k}|^{2}\left(|M_*|^{2}-\frac{\mu|M_*|^{2}|M_{k-1}|^{2}}{\nu_1+|M_{k-1}|^2}\right)}{(1+a|y_{k}|^{2})(\nu_1+|M_{k-1}|^2)}\nonumber\\
&\qquad+\frac{a\mu|y_{k}|^{2}|\tilde M_{k-1}|^{2}}{(1+a|y_{k}|^{2})(\nu_1+|M_{k-1}|^2)}\nn\\
&\qquad-b\gamma\frac{|M_{k-1}|^2|y_{k}|^2|\tilde M_{k-1}|^{2}}{\nu_2+|M_{k-1}|^2|y_{k}|^2}.
\end{align}
Since for all $k\in\BBZ^+$, $\Delta V_M(k)\leq 0$, it follows that $|M_{k-1}|\leq |\tilde M_{k-1}|+|M_*|\leq |\tilde M_{0}|+|M_*|\leq |M_{0}|+2|M_*|$, and thus $|M_{k-1}|^2\leq a\nu_2$. Therefore, since $\mu\in(0,1]$, it follows from \eqref{Eq:ineq20} that
\begin{align}
\label{Ineq45}
\Delta V(k)&\leq-\frac{c_1|y_{k}|^{2}}{1+a|y_{k}|^{2}}+\frac{a\mu|y_{k}|^{2}|\tilde M_{k-1}|^{2}}{\nu_1(1+a|y_{k}|^{2})}\nn\\
&\qquad-b\gamma\frac{|M_{k-1}|^2|y_{k}|^2|\tilde M_{k-1}|^{2}}{\nu_2+a\nu_2|y_{k}|^2},
\end{align}
where $c_1\triangleq \frac{a\mu\nu_1|M_*|^2}{(\nu_1+a\nu_2)^2}>0$.

To show that $(0,M_*)$ is a Lyapunov stable equilibrium, define $\SD\triangleq \BBC\times\left\{x\in\BBC:|x|<|M_*|/2\right\}$, and note that for all $(y_{k},\tilde M_{k-1})\in\SD$, $|M_{k-1}|\geq |M_*|/2$. Let $b\triangleq \frac{4a\mu\nu_2}{\gamma\nu_1 |M_*|^2}$, and it follows from \eqref{Ineq45} that for all $(y_{k},\tilde M_{k-1})\in\SD$, $\Delta V(k)\leq -c_1|y_{k}|^{2}/(1+a|y_{k}|^{2})$, which is nonpositive. Therefore, $(y_{k},M_{k-1})\equiv(0,M_*)$ is a Lyapunov stable equilibrium.

To show convergence of $y_k$ and $u_k$, let $M_0\in\SM\backslash\{0\}$, and define
\begin{align}
\label{c2}
c_2\triangleq\begin{cases} |M_*|,&{\rm if~}M_0=M_*,\\|{\rm Im~}\tilde M_0M_*^*|/|\tilde M_0|,&{\rm if~}M_0\neq M_*.\end{cases}
\end{align}
First, assume $M_0=M_*$, and it follows that for all $k\in\BBN$, $M_k=M_*$. In this case, for all $k\in\BBN$, $|M_k|=c_2$.
 
Next, assume $M_0\neq M_*$. Since $M_0\in\SM\backslash\{0\}$, it follows that, $\angle M_*-\angle M_0\neq \pi$, which implies that $\angle( M_0/M_*)\neq\pi$. Thus, ${\rm Im}(M_0/M_*)\neq 0$, which implies that ${\rm Im~}\tilde M_0M_*^*={\rm Im~}(\tilde M_0M_*^*+M_*M_*^*)={\rm Im~}M_0M_*^*=|M_*|^2{\rm Im}(M_0/M_*)\neq 0$,
and it follows from \eqref{c2} that $c_2>0$. Next, note that it follows from \eqref{Eq:closed-loop-siso} and \eqref{Eq:c_updateSISO} that
\begin{align*}
\tilde{M}_{k}&=\left(1-\dfrac{\gamma |M_{k-1}|^2|y_{k}|^2}{\nu_2+|M_{k-1}|^2|y_{k}|^2}\right)\tilde {M}_{k-1},
\end{align*}
which has the solution $\tilde M_{k}=\beta_k\tilde M_{0}$,
where $\beta_{k}\triangleq \prod_{i=0}^{k-1}\left(1-\dfrac{\gamma |M_i|^2|y_{i+1}|^2}{\nu_2+|M_{i}|^2|y_{i+1}|^2}\right)$. Thus, for all $k\in\BBZ^+$, $|M_{k}|^{2}=|\tilde M_{k}+M_*|^{2}=|\tilde M_{0}\beta_{k}+M_*|^{2}=|\tilde M_{0}|^{2}\beta_{k}^{2}+2({\rm Re~}\tilde M_{0} M_*^*)\beta_{k}+|M_*|^{2}.$
Since $f(x)\triangleq |\tilde M_{0}|^{2}x^{2}+2({\rm Re~}\tilde M_{0} M_*^*)x+|M_*|^{2}$ is quadratic and positive definite in $x$, it follows that $f$ is minimized at $\frac{-{\rm Re~}\tilde M_0M_*^*}{|\tilde M_0|^2}$. Thus, for all $k\in\BBN$,
\begin{align*}
|M_{k}|^2&\geq \frac{|M_*|^2|\tilde M_0|^2-\left({\rm Re~}\tilde M_0M_*^*\right)^2}{|\tilde M_0|^2}=c_2^2.
\end{align*}
Thus, for all $k\in\BBN$, $|M_k|>c_2$. Let $b\triangleq \frac{ a\mu\nu_2}{\gamma\nu_1 c_2^2}$, and it follows from \eqref{Ineq45} that
\begin{align}
\label{DeltaV_main}
\Delta V(k)&\leq-\frac{c_1|y_{k}|^{2}}{1+a|y_{k}|^{2}}.
\end{align}
Next, since $V$ is positive-definite, and for all $k\in\BBN$, $\Delta V(k)$ is nonpositive, it follows from \eqref{DeltaVmaa} and \eqref{DeltaV_main} that $0\leq \lim_{k\to\infty}\sum_{i=1}^{k}\frac{c_1|y_{i}|^{2}}{1+a|y_{i}|^{2}}\leq-\lim_{k\to\infty}\sum_{i=1}^{k}\Delta V(i)=V(y_{1},\tilde M_{0})-\lim_{k\to\infty} V(y_{k},\tilde M_{k-1})\leq V(y_{1},\tilde M_{0}),$
where the upper and lower bounds imply that all the limits exist. Thus,
 $\lim_{k\to\infty}\frac{c_1|y_{k}|^{2}}{1+a|y_{k}|^{2}}=0$. Since in addition, $\frac{c_1|y_{k}|^{2}}{1+a|y_{k}|^{2}}$ is a positive-definite function of $y_{k}$, it follows that $\lim_{k\to\infty} y_{k}=0$. Furthermore, \eqref{Eq:open-loop-MIMO} implies that $\lim_{k\to\infty}u_k=\lim_{k\to\infty}(y_{k+1}-\hat d)/M_*=u_*$.
\end{proof}
\bibliographystyle{unsrt}
\bibliography{MasterReferenceList}
\end{document}